\documentclass[showpacs,reprint,twocolumn,preprintnumbers,amsmath,amssymb,aps,superscriptaddress]{revtex4-1}

\usepackage{graphicx}
\usepackage{dcolumn}
\usepackage{bm}
\usepackage{subfigure}
\usepackage{color}
\usepackage{times}

\renewcommand{\eqref}[1]{(\ref{#1})}

\begin{document}
\title{Quantum Fisher Information as a Predictor of Decoherence in the Preparation of Spin-Cat States for Quantum Metrology}
\author{Samuel~P.~Nolan}
\email{uqsnolan@uq.edu.au}
\affiliation{School of Mathematics and Physics, The University of Queensland, Brisbane, Queensland, Australia}
\author{Simon~A.~Haine}
\affiliation{Department of Physics and Astronomy, University of Sussex, Brighton, United Kingdom}

\date{\today}

\begin{abstract}

In its simplest form, decoherence occurs when a quantum state is entangled with a second state, but the results of measurements made on the second state are not accessible. As the second state has effectively ``measured" the first, in this paper we argue that the quantum Fisher information is the relevant metric for predicting and quantifying this kind of decoherence. The quantum Fisher information is usually used to determine an upper bound on how precisely measurements on a state can be used to estimate a classical parameter, and as such it is an important resource. Quantum enhanced metrology aims to create non-classical states with large quantum Fisher information and utilise them in precision measurements. In the process of doing this it is possible for states to undergo decoherence, for instance atom-light interactions used to create coherent superpositions of atomic states may result in atom-light entanglement. Highly non-classical states, such as spin-cat states (Schr{\"o}dinger cat states constructed from superpositions of collective spins) are shown to be highly susceptible to this kind of decoherence. We also investigate the required field occupation of the second state, such that this decoherence is negligible.

\end{abstract}

\pacs{37.25.+k, 03.75.Gg}

\maketitle

\section{Introduction}\label{sec1} 

%






Quantum metrology is the science exploiting quantum correlations to estimate a classical parameter $\chi$, such as a phase, beyond the sensitivity available in uncorrelated systems. Given a metrological scheme with access to $N$ total particles there is an upper bound on the precision, $\Delta \chi \geq 1/N$, called the Heisenberg limit \cite{Holland1993, Giovannetti2006}. For a two-mode interferometer with conserved total particle number, called an $SU(2)$ interferometer, the class of states which yield Heisenberg limited sensitivity are spin-cat states, an example of which is the well known $NOON$ state, which achieves Heisenberg limited sensitivities via a parity measurement \cite{Bollinger1996, Dowling2002, Leibfried2004, Giovannetti2004}.

As $NOON$ states are highly non-classical, possibly massive superpositions, they could also find a number of applications outside quantum metrology. In particular these states could be well suited for testing macroscopic realism \cite{Leggett1985, Arndt2014}, gravitational decoherence \cite{Derakhshani2016}, spontaneous wavefunction collapse theories \cite{Bassi2003, Adler2009, Bassi2013} as well as realising the Greenberger, Horne and Zeilinger (GHZ) state, which could test local hidden variable theories \cite{Greenberger1990}. Optical GHZ states could also find applications in quantum communication and computation \cite{Bose1998, Hillery1999, Panangaden2005, Qin2007}.

A spin-cat state in a Bose-Einstein condensate would be well suited to a number of these applications, particularly metrology. However, this state has yet to be realised, due to the immense challenge of maintaining the quantum coherence of the state \cite{Aolita2008, DemkowiczDobrzanski2012}. This is despite a number of proposed methods, previously relying on Josephson coupling between two modes \cite{Savage1999, Cirac1998, Weiss2007}, collisions of bright solitons \cite{Weiss2009}, and through the atomic Kerr effect \cite{Dunningham2001, Dunningham2006a, Dunningham2006b, Lau2014}. Although the atomic interaction times have been too small to generate spin-cat states, the Kerr effect has successfully been used to generate large numbers of entangled particles in Bose-Einstein condensates \cite{Esteve2008, Riedel2010, Gross2010, Leroux2010}. There has also been success outside the realm of quantum atom-optics, with $NOON$ states having been realised in modestly sized systems such as superconducting flux qubits \cite{Friedman2000}, optics \cite{Grangier2007}, and in trapped ions \cite{Monroe1996, Liebfried2005, Monz2011}, the latter with up to 14 particles.

In any case, to actually do anything useful with such a state, it may be necessary to perform a unitary rotation. This could be, for example, to prepare the state for input into an interferometer. However, unitary evolution is only an approximation, valid when the system used to perform this operation is sufficiently large such that it can be considered \emph{classical}. 

In this paper we relax this approximation, and investigate rotations caused by interaction with a quantized auxiliary system. As an example, consider a two-component atomic Bose-Einstein condensate. A rotation of the state on the Bloch-sphere can be implemented by interaction with an optical field via the AC Stark shift \cite{Scullybook}. It's often the case that the number of photons in this state is sufficiently large that the quantum degrees of freedom of the light are ignored. However, in metrology we are are often interested in quantum states that are particularly sensitive to decoherence, such as spin-cat states, therefore in this paper we investigate the effect of treating this optical field as a quantized auxiliary system. Decoherence in systems such as these has been considered previously \cite{Dalibard1992}, using a stochastic wavefunction approach in small systems. Although QFI of quantum states with decoherence has also been considered in the literature \cite{Zhong2013, Huang2015a, Altintas2016}, the goal of this paper is to employ new approach, by defining a quantum Fisher information for the optical field.

It has been shown that in the presence of entanglement between the state and some auxiliary system, the metrological usefulness of the state may be enhanced by allowing measurements on the auxiliary system \cite{Hammerer2010, Haine2013, Szigeti2014, Tonekaboni2015, Haine2015, Haine2015a, Haine2016b}. However in this paper we take a different approach, and study the metrological usefulness of a state if measurements of the auxiliary system are forbidden. The goal is not to devise schemes to enhance metrological sensitivity, but to study the sensitivity of quantum states (particulary spin-cat states) to this kind of decoherence.

After introducing the formalism in which we work in Section \ref{sec2}, we demonstrate the central idea of this paper in Section \ref{sec3} by studying the intuitive case of a simple operator product Hamiltonian. In this situation a number of results may be obtained analytically, which we use to understand decoherence in terms of the noise properties of the initial auxiliary state. In section \ref{sec4} we turn our attention to a beam-splitter Hamiltonian, which is less intuitive. In section \ref{sec5} we introduce a semi-classical formalism which gives us a simple picture of this decoherence, and also an efficient means of simulating the full composite system. Finally in section \ref{sec6} we apply this method to study the limits of this decoherence, deriving the required auxiliary field occupation to negate significant entanglement between the systems.

\section{Formalism} \label{sec2}

The generic problem considered in quantum metrology is this: given an initial quantum state $\hat{\rho}(0)$ that undergoes unitary evolution $\hat{U}_\chi=\exp( -i \hat{G} \chi)$, how precisely can the classical parameter $\chi$ be estimated? The answer is given by the quantum Cram{\'e}r-Rao bound (QCRB) which places a lower bound on the sensitivity, i.e. $\Delta \chi \geq 1/\sqrt{\mathcal{F}}$, where
\begin{equation} \label{eq:QFI}
\mathcal{F} = 2 \sum_{i,j} \frac{(\lambda_i-\lambda_j)^2}{\lambda_i+\lambda_j} | \langle e_i| \hat{G} |e_j \rangle |^2
\end{equation}
is the quantum Fisher information (QFI) \cite{Caves1994, Paris2009, Toth2014, DemkowiczDobrzanski2015} and $\lambda_i$, $|e_i \rangle$ are the eigenvalues and eigenvectors of $\hat{\rho}(0)$. When $\hat{\rho}(0)$ is pure, Eq.~\eqref{eq:QFI} reduces to the variance of $\hat{G}$, specifically $\mathcal{F} = 4 V( \hat{G})$. The QFI does not depend on the choice of a particular measurement signal, only on the input state and the Hermitian operator $\hat{G}$, called the generator of $\chi$. 

\begin{figure}
\includegraphics[width=\columnwidth]{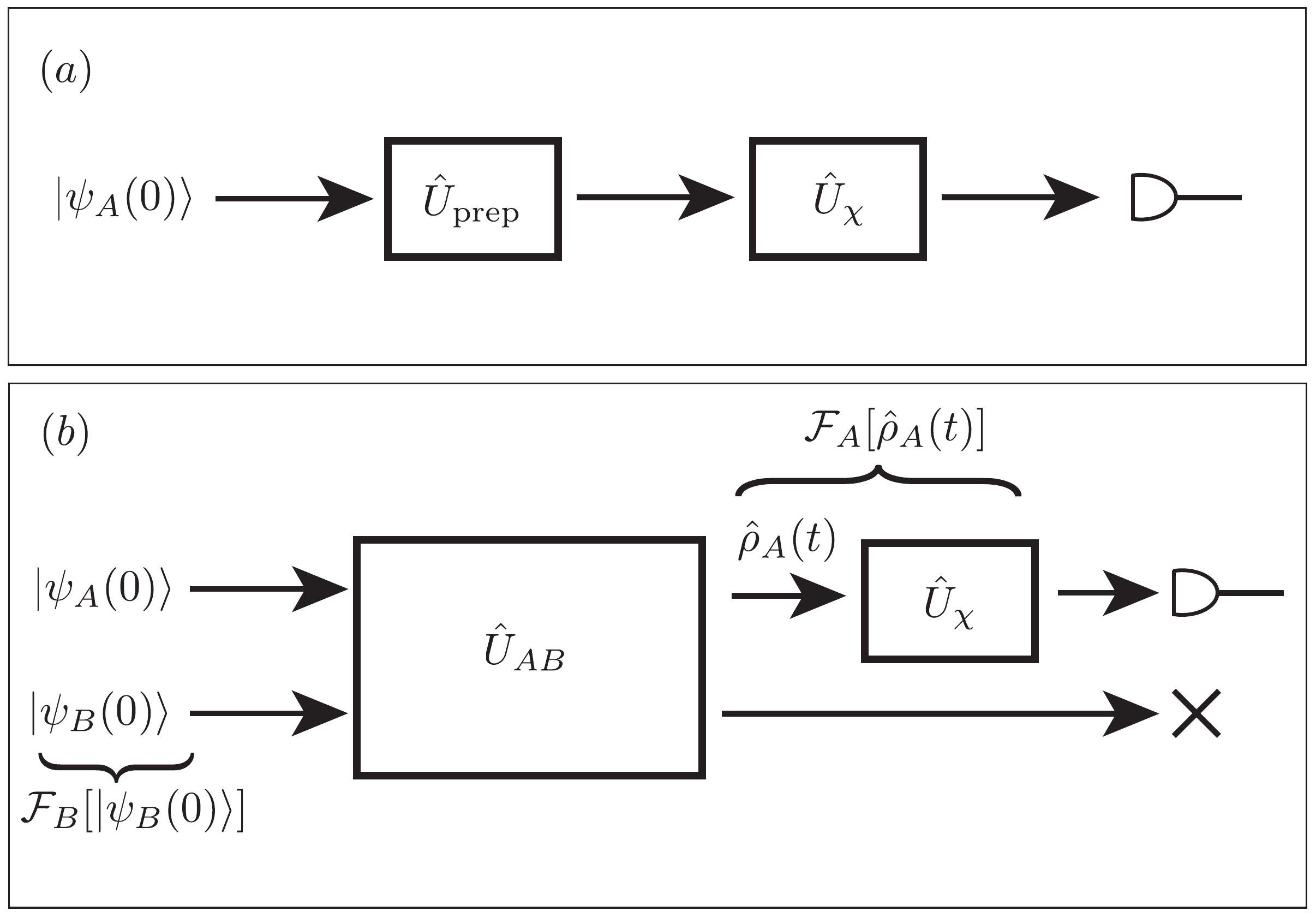} 
\caption{A visual summary of our scheme. (a) The ideal situation in quantum metrology. A state $|\psi_A(0) \rangle$ is prepared unitarily for input into the interferometer $\hat{U}_\chi=\exp( -i \hat{G}_A \chi)$, i.e. the quantum Fisher information of $\hat{U}_{\rm prep} |\psi_A(0) \rangle$ with respect to $\hat{G}_A$ is valuable. (b) If the preparation involves interaction with some other quantum system, then they may become entangled. If we cannot measure anything about system $B$ then system $A$ is now mixed, and we have lost quantum Fisher information with respect to $\hat{U}_{\mathrm{prep}}|\psi_A (0)\rangle$. In this paper we investigate this decoherence, and determine under what circumstances scheme (b) is well approximated by scheme (a).}
\label{fig:diagram}
\end{figure}

The system we consider in this paper is illustrated in Fig.~\ref{fig:diagram}. We begin with some quantum state $|\psi_A\rangle$, called the probe state, which could be used to probe a classical parameter $\chi$. Before this happens the state must be prepared in some way. Ideally, this would occur by performing some unitary operation $\hat{U}_\mathrm{prep}$ on the initial state $|\psi_A(0)\rangle$: $|\psi_A\rangle = \hat{U}_\mathrm{prep}|\psi_A(0)\rangle$ [Fig.~\ref{fig:diagram}(a)]. However in practice, treating this preparation step as unitary is usually an approximation, as the physical mechanism to achieve this preparation can involve entanglement with some auxiliary subsystem $B$. In this case we replace $\hat{U}_\mathrm{prep}$ (that is assumed to operate only on subspace $A$) with $\hat{U}_{AB} = \exp (-i \hat{\mathcal{H}}_{AB}t/\hbar)$, which can potentially cause entanglement between subsystems $A$ and $B$, and therefore cause decoherence in subsystem $A$ when system $B$ is ignored. In this case system $A$ is described by the state $\hat{\rho}_A = \rm{Tr}_B \{|\psi_{AB}\rangle\langle\psi_{AB}| \}$, where $|\psi_{AB}\rangle = \hat{U}_{AB}|\psi_A(0)\rangle \otimes |\psi_B(0)\rangle$ [Fig.~\ref{fig:diagram}(b)].  

To illustrate this concept, consider the example of an optical parametric oscillator (OPO), used to create the well known squeezed vacuum states by creating pairs of photons via a Hamiltonian $\hat{\mathcal{H}}_A = \eta\left(\hat{a}\hat{a} + \hat{a}^\dag\hat{a}^\dag\right)$  \cite{Scullybook}. The physical mechanism that achieves this process involves the annihilation of a photon of twice the frequency from a \emph{pump} beam, which we label system $B$. This process is described by the Hamiltonian $\hat{\mathcal{H}}_{AB} =  g\left(\hat{b}^\dag\hat{a}\hat{a} + \hat{b}\hat{a}^\dag\hat{a}^\dag\right)$. In this context, the approximation that the entanglement between the systems can be ignored such that $\hat{\mathcal{H}}_{AB}$ can be replaced with $\hat{\mathcal{H}}_A$ is often referred to as the \emph{undepleted pump approximation}. While this is usually a good approximation, there are experimentally accessible regimes where it becomes invalid \cite{Kherunstyan2005, Olsen2006, Lewis-Swan2013}. If we do not permit measurements on $B$, then the entanglement between the two systems will result in decoherence, which we quantify as a reduction in the QFI of the probe system $\mathcal{F}_A$, because $\mathcal{F}_A \leq 4 V( \hat{G}_A)$. 

In what follows we work in the  standard formalism for $SU(2)$ interferometers \cite{Yurke1986}, whereby our probe system consists a conserved total number of $N_A=\langle \hat{a}_1^\dagger \hat{a}_1 + \hat{a}_2^\dagger \hat{a}_2\rangle$ bosons each in one of two modes, with bosonic annihilation operators $\hat{a}_1$ and $\hat{a}_2$ respectively. Collective observables are represented as pseudo-spin operators $\hat{J}_k = \frac{1}{2}(\hat{a}_1^\dagger \ \hat{a}_2^\dagger) \sigma_k (\hat{a}_1 \ \hat{a}_2)^T$, where $\sigma_k$ is the $k$th Pauli matrix, hence the system is described by the well known $SU(2)$ algebra. The auxiliary system has a mean field occupation of $\langle \hat{n}_B \rangle = N_B$ bosons, which for simplicity are confined to a single mode $\hat{b}$, with number operator $\hat{n}_B=\hat{b}^\dagger \hat{b}$. 

In the absence of quantum correlations between particles, an ensemble of two-level particles is well described by a coherent spin-state (CSS) \cite{Arecchi1972,Radcliffe1971}, which can be thought of as a rotation of the maximal $J_z$ eigenstate: $|CSS \rangle = |\theta, \phi \rangle = \hat{R}(\theta, \phi)|N_A,0 \rangle$ where (up to a global phase) $\hat{R}(\theta, \phi)$ is the rotation operator,
\begin{equation} \label{eq:Rotop}
\hat{R}(\theta, \phi)= e^{-i\hat{J}_z  \phi } e^{-i \hat{J}_x \theta},
\end{equation}
and the state $|N_1, N_2\rangle$ indicates $N_{1(2)}$ bosons in mode $1$(2). Coherent spin-states have the useful property that they are an extreme eigenstate of the rotated pseudo-spin operator $\hat{J}_{\theta,\phi}=\hat{R}^\dagger (\theta, \phi) \hat{J}_z \hat{R} (\theta, \phi)=\hat{J}_z \cos(\theta)+[\hat{J}_x \sin(\phi)+\hat{J}_y \cos(\phi)] \sin(\theta)$. 

We are interested in a class of states called spin-cat (SC) states, which are an equal superposition of opposite coherent spin-states, i.e. the maximum and minimum $\hat{J}_{\theta, \phi}$ eigenstates,
\begin{equation} \label{eq:catdfn}
|SC\rangle =\frac{1}{\sqrt{2}}( | \rm{max} \rangle+e^{i \vartheta} | \rm{min} \rangle ).
\end{equation}
These states are highly non-classical, and have the maximum QFI for an $SU(2)$ interferometer, $\mathcal{F}_A=N_A^2$, so long as $\hat{G}_A=\hat{J}_{\theta, \phi}$. In contrast, coherent spin-states are shot-noise limited, with $\mathcal{F}_A\leq N_A$ with respect to $\hat{G}_A=\hat{J}_{\theta, \phi}$.

When $\hat{J}_{\theta, \phi}=\hat{J}_z$, the spin-cat state is the well known $NOON$ state, $|NOON \rangle = (|N_A,0 \rangle + |0, N_A \rangle)/\sqrt{2}$ \cite{Sanders1989, Boto2000}. $NOON$ states are particularly relevant as many experiments would be limited to performing measurements on the probe system in the number basis. However another relevant basis is $\hat{J}_y$, as it is straight forward to show that the well known one axis twisting interaction will eventually lead to a spin-cat in the $\hat{J}_y$ basis, i.e. $|SC \rangle = \exp(-i \hat{J}_z^2 \pi/2)|\pi/2,0 \rangle$.

\section{Separable Interactions} \label{sec3}
In this section we consider the case where the interaction Hamiltonian between systems $A$ and $B$ is a separable tensor product of operators acting on each Hilbert space. Specifically,\
\begin{equation}
\hat{\mathcal{H}}_{AB} = \hbar g \hat{H}_A\otimes \hat{G}_B \, . \label{eq:sepham}
\end{equation}
Such an interaction may arise when the Hermitian operator $\hat{H}_A$ is required in the state preparation of system $A$, but is moderated by the Hermitian operator $\hat{G}_B$ acting on subspace $B$.

\subsection{Some General Results}\label{sec3a}
For an initially separable and pure state $|\psi_{AB}(0) \rangle = |\psi_A(0) \rangle \otimes | \psi_B(0) \rangle$, in terms of the dimensionless time $\tau=gt$ the evolved state is
\begin{equation} \label{eq:psifull}
| \psi_{AB}(\tau) \rangle = \sum_{m} c_m |m \rangle \otimes \left(e^{-i \lambda_m \hat{G}_B \tau} | \psi_B(0) \rangle \right) \, .
\end{equation}
In terms of the eigenstates $|m\rangle$ and eigenvalues $\lambda_m$ of $\hat{H}_A$, the reduced density matrix $\hat{\rho}_A=\mathrm{Tr}_B\{| \psi_{AB}(\tau) \rangle \langle \psi_{AB}(\tau) | \}$ takes the simple form 
\begin{equation} \label{eq:rhoA}
\hat{\rho}_A(\tau) = \sum_{m,n}c_m c^*_n \mathcal{C}_{m,n}(\tau) |m \rangle \langle n | \, ,
\end{equation}
where $\langle j | \psi_A(0) \rangle=c_j$ is the initial state. We have defined 
\begin{equation} \label{eq:coh}
\mathcal{C}_{m,n}(\tau) = \langle \psi_B(0) |e^{-i \left(\lambda_m-\lambda_n \right) \hat{G}_B \tau} | \psi_B(0) \rangle,
\end{equation}
which we call the coherence matrix of the probe system, as it is responsible for the decay of the off-diagonal terms of $\rho_A$. This term is a direct consequence of a partial trace over the auxiliary system $B$.

If $\mathcal{C}_{m,n}=1$ then $\hat{\rho}_A$ remains pure, and if $\mathcal{C}_{m,n}=\delta_{m,n}$ then $\hat{\rho}_A$ is a completely incoherent mixture of eigenstates $|m \rangle \langle m |$. More generally the relationship between the purity of the probe system $\gamma = \rm{Tr} \{(\hat{\rho}_A)^2 \}$ and $\mathcal{C}_{m,n}$ is
\begin{equation} \label{eq:purity}
\gamma = \sum_{m,n} |c_{m}|^2 |c_{n}|^2 |\mathcal{C}_{m,n}|^2.
\end{equation}

As we are interested in maintaining states with high values of $\mathcal{F}_A$, we are particularly interested in the magnitude of $|\mathcal{C}_{m,n}|^2$, as states with lower purity usually have reduced QFI. Expanding the magnitude of $\mathcal{C}_{m,n}$ to second order in even powers of $\Delta_{m,n} = (\lambda_m-\lambda_n) \tau$ (odd powers do not contribute) reveals a link between the QFI of the auxiliary system with respect to $\hat{G}_B$ and the resultant decoherence in the probe system:
\begin{equation} \label{eq:cohexpand}
|\mathcal{C}_{m,n}|^2 = 1-\frac{\mathcal{F}_B}{4} \Delta_{m,n}^2 + \mathcal{O}(\Delta_{m,n}^4),
\end{equation}
where $\mathcal{F}_B = 4 V(\hat{G}_B)$ is the QFI of $|\psi_B(0)\rangle$ associated with measuring some classical parameter $\eta$ under evolution $\hat{U}_{\eta}=\exp(-i \hat{G}_B \eta)$. For short times at least, we identify this QFI as being the relevant parameter to predict the decay of the off-diagonal matrix elements of $\hat{\rho}_A$.  

Such an identification is particularly intuitive for considering the role of $|\psi_B(0)\rangle$ in the decoherence of system $A$: if one considers the possibility that the outgoing state of system $B$ could be \emph{measured} by an observer, then if this state carries information which can distinguish between the eigenvalues $\lambda_m$ and $\lambda_n$, we no longer expect there to be a coherent superposition of these components. The interaction with system $B$ effectively \emph{measured} system $A$.  That is, states with high QFI with respect to their ability to estimate the physical observable corresponding to $\hat{H}_A$ cause the most rapid decoherence. 

Even if $\mathcal{C}_{m,n}$ is known, calculating the QFI of the probe system requires diagonalization of the reduced density matrix [see Eq.~\eqref{eq:QFI}]. Fortunately for evolution under a separable Hamiltonian, some simple analytic results exist for some initial states. For any state that is initially a spin-cat state of extreme $\hat{J}_{\theta, \phi}=\hat{H}_A$ eigenstates, there is a simple relationship between the probe QFI and the purity of the reduced density matrix:
\begin{equation} \label{eq:analNoonQFI}
\mathcal{F}_A=N_A^2 \left(2 \gamma -1 \right) ,
\end{equation}
and from Eq.~\eqref{eq:purity} the purity is given by
\begin{equation} \label{eq:analNoonpurity}
\gamma = \frac{1}{2} \left(1+ |\mathcal{C}_{\rm max}|^2 \right) ,
\end{equation}
where $\mathcal{C}_{\rm max}=\mathcal{C}_{N_A+1, 1}=(\mathcal{C}_{1, N_A+1})^*$ is the extreme off-diagonal term of the coherence matrix, i.e. for a spin-cat in $\hat{J}_{\theta,\phi}$, the QFI of the auxiliary system depends only on the purity of $\hat{\rho}_A$, which at least for short times depends only on the QFI of the initial auxiliary state $|\psi_B (0)\rangle$, i.e. $\mathcal{F}_A$ is a function only of $\mathcal{F}_B$ and time.

From these relations, to second order in $|\mathcal{C}_{\mathrm{max}}|^2$ [Eq.~\eqref{eq:cohexpand} with $\lambda_m-\lambda_n=N_A$] we have
\begin{equation} \label{eq:purityapprox}
\gamma \approx \exp \left( -\frac{1}{8} \mathcal{F}_B N_A^2 \tau^2\right)
\end{equation}
and
\begin{equation} \label{eq:FAapprox}
\mathcal{F}_A \approx N_A^2 \exp \left( - \frac{1}{4} \mathcal{F}_B N_A^2 \tau^2 \right) .
\end{equation}
Although these relations only hold for small time, they do not assume anything about the input state of the auxiliary system. For any $|\psi_B(0)\rangle$, the QFI with respect to $\hat{G}_A=\hat{H}_A$ and purity of a $\hat{J}_{\theta, \phi}=\hat{H}_A$ spin-cat state simply decay exponentially with $\mathcal{F}_B$ and time squared, at a rate proportional to $N_A^2$, as one might expect for a state capable of reaching the Heisenberg limit. This kind of scaling has been seen in previous studies of Heisenberg limited states under decoherence \cite{Aolita2008, DemkowiczDobrzanski2012}, but not in this context.

\subsection{An Example} \label{sec3b}

We will now study the decoherence imparted on a probe after evolution under a $\hat{H}_A=\hat{J}_z$ rotation, specifically
\begin{equation} \label{eq:Jzrot}
\hat{\mathcal{H}}_{AB} = \hbar g \hat{J}_z \otimes \hat{n}_B.
\end{equation}
This kind of interaction describes a number of systems, for instance superconducting qubits coupled to a microwave cavity \cite{Wallraff2004, Schuster2005, Haigh2015}, or the weak probing of an ensemble of two-level atoms with light detuned far from resonance \cite{Szigeti2009, Hammerer2010, Wasilewski2010, Chen2011, Szigeti2010, Leroux2010, Vanderbruggen2011, Bernon2011, Brahms2012, Bohnet2014, Haine2015a}. This Hamiltonian generates a $\hat{J}_z$ rotation, which corresponds to a relative phase being imparted between the two levels available to the probe system. Although we are agnostic about the specific system being studied, for convenience we will adopt the language of atom-light interactions, and will often refer to the quanta of the auxiliary field as photons.

\begin{figure*} 
\includegraphics[width=\textwidth]{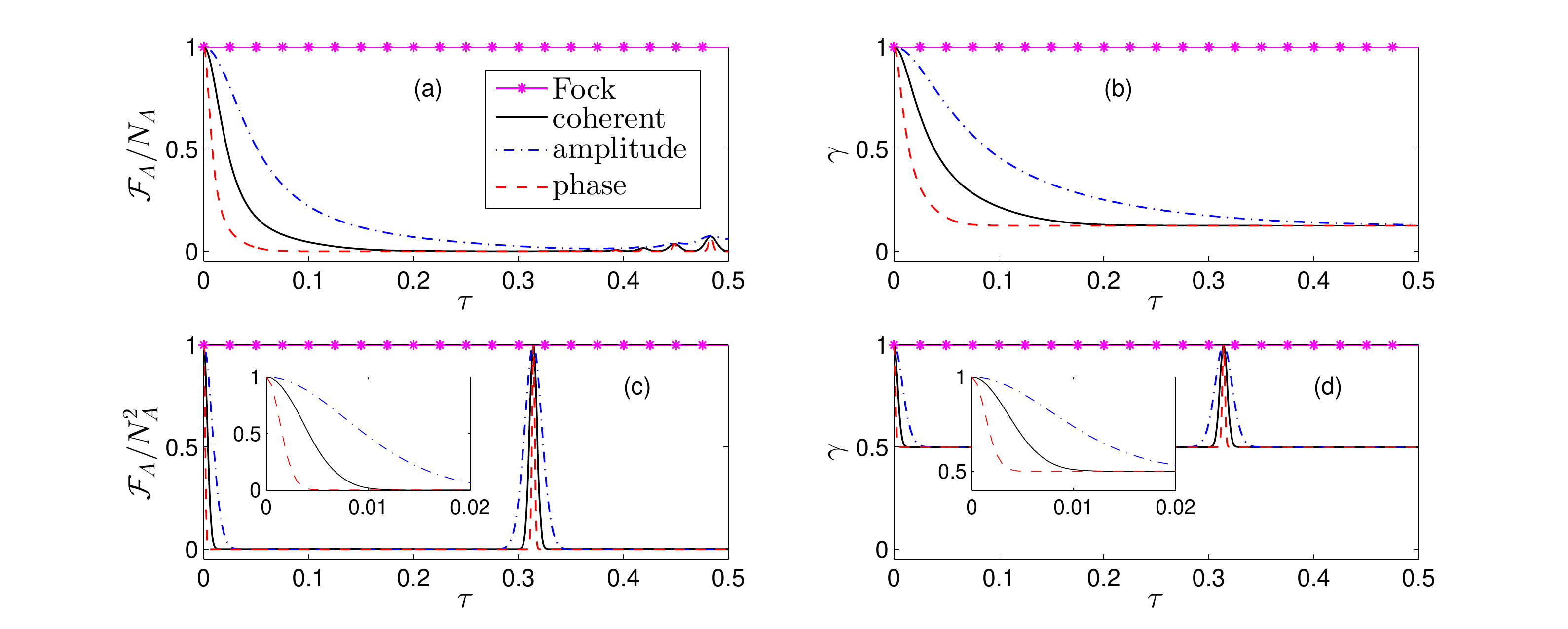}
\caption{(Color online). Time evolution under $\hat{\mathcal{H}}_{AB} = \hbar g \hat{J}_z \otimes \hat{n}_B$ of the quantum Fisher information $\mathcal{F}_A$ and purity $\gamma$ of $\hat{\rho}_A(\tau)$, for a variety of initial states with average particle number $N_A=20$ and $N_B=100$ and dimensionless time parameter $\tau=gt$. Plots (a),(b) have initial CSS $|\psi_A(0) \rangle =|\pi/2,0\rangle$ and (c), (d) have $|\psi_A(0) \rangle = |NOON \rangle$. Note that the normalization of $\mathcal{F}_A$ is different for (a) compared to (c). Insets in (c), (d) show short evolution times. Each plot also varies the initial auxiliary state for the cases discussed in the text, i.e. a Glauber-coherent state (solid black), an amplitude squeezed state with $r=1$ (dashed blue), a phase squeezed state with $r=-1$ (dashed red) and a Fock state (magenta asterisks); the Fock states have been included for completeness, although they have $\mathcal{F}_B=0$ so no coherence is lost. For all auxiliary input states we chose $\beta$ to be real. Time scales can vary greatly between experimental systems, for instance $g\approx 10^2$ rad/s \cite{Lee2014}, $10^4$ rad/s \cite{Zhang2012}, up to values as high as $10^6$ rad/s \cite{Brennecke2007}.}
\label{fig:varyt}
\end{figure*}

An interaction of the form Eq.~\eqref{eq:Jzrot} leads to entanglement between the $J_z$ spin projection of system $A$ and the the phase of system $B$, as $\hat{n}_B$ is the generator of phase. Identifying $\hat{G}_B=\hat{n}_B$, it is immediately obvious that the optimal choice for $|\psi_B(0) \rangle$ is a Fock state, (i.e. an $\hat{n}_B$ eigenstate) as this state has $\mathcal{F}_B=4 V(\hat{n}_B) = 0$, and the operation can be performed without generating any entanglement between the systems, i.e. $|\mathcal{C}_{m,n}|^2=1$, which is illustrated in Fig.~\ref{fig:varyt}. This is consistent with our view of system $B$ carrying away information about $J_z$, as Fock states have entirely undefined phase, so cannot be used to make a measurement via the interaction Eq.~\eqref{eq:Jzrot}. However, as Fock states are difficult to engineer, it is important to consider the behaviour of other states. 
 
Throughout this paper we will focus on commonly accessible states such as Glauber coherent states and quadrature squeezed states, which have the form
\begin{equation} \label{eq:psiB}
|\psi_B(0) \rangle = \hat{D}(\beta) \hat{S}(r) | 0 \rangle \, ,
\end{equation}
where $\hat{D}(\beta)=\exp(\beta \hat{b}^\dagger-\beta^* \hat{b})$ is the coherent displacement operator with coherent amplitude $\beta$, $\hat{S} = \exp[r( \hat{b}^2-(\hat{b}^\dagger)^2)/2]$ is the single mode squeezing operator with real squeezing parameter $r$ and optical vacuum $|0 \rangle$. In particular we focus on three cases, the Glauber-coherent state ($r=0$), the amplitude squeezed state ($r>0$) and the phase squeezed state ($r<0$), which, for a fixed mean photon number, have $\mathcal{F}_B^\mathrm{phase}>\mathcal{F}_B^\mathrm{coherent}>\mathcal{F}_B^\mathrm{amplitude}$. 

It is possible to evaluate the coherence matrix [Eq.~\eqref{eq:coh}] analytically for these states. Because we have $\hat{H}_A=\hat{J}_z$ the spin-cat states that obey the relations Eq.~\eqref{eq:analNoonQFI} and Eq.~\eqref{eq:analNoonpurity} are $NOON$ states. The generator $\hat{G}_A=\hat{J}_z$ has integer eigenvalues, and so we make the substitution $\lambda_m-\lambda_n = m-n$. For simplicity we will restrict ourselves to real $\beta$, although it is not necessary to do so. By observing that $\exp[-i(m-n) \hat{n}_B \tau] \hat{D}(\beta) \hat{S}(r) |0 \rangle = \hat{D}(\beta') \hat{S}(r') |0 \rangle$, where $\beta'=\beta \exp[-i(m-n)\tau]$ and $r'=r \exp[-2i(m-n)\tau]$, the problem is reduced to evaluating the overlap of two squeezed coherent states, see for instance \cite{Yuen1976,Caves1985}. We obtain
\begin{equation} \label{eq:cohsqueeze}
\mathcal{C}_{m-n}(\beta, r, \tau) = \frac{\exp \left({\frac{\beta^2 \left[1+\coth(r) \right] \left[ e^{-i(m-n)\tau}-1\right]}{e^{-i(m-n) \tau}+\coth(r)}} \right)}{\sqrt{\cosh^2(r)-e^{-2i (m-n) \tau} \sinh^2(r)}},
\end{equation}
as the coherence matrix for a squeezed coherent state. For completeness we also provide the result for a Glauber-coherent state, obtained by simply taking the limit $r \rightarrow 0$,
\begin{equation} \label{eq:cohbeta}
\mathcal{C}_{m-n}(\beta,\tau)=\exp\left[ N_B \left(e^{-i  (m-n)\tau} -1 \right) \right].
\end{equation}

\begin{figure}
\includegraphics[width=\columnwidth]{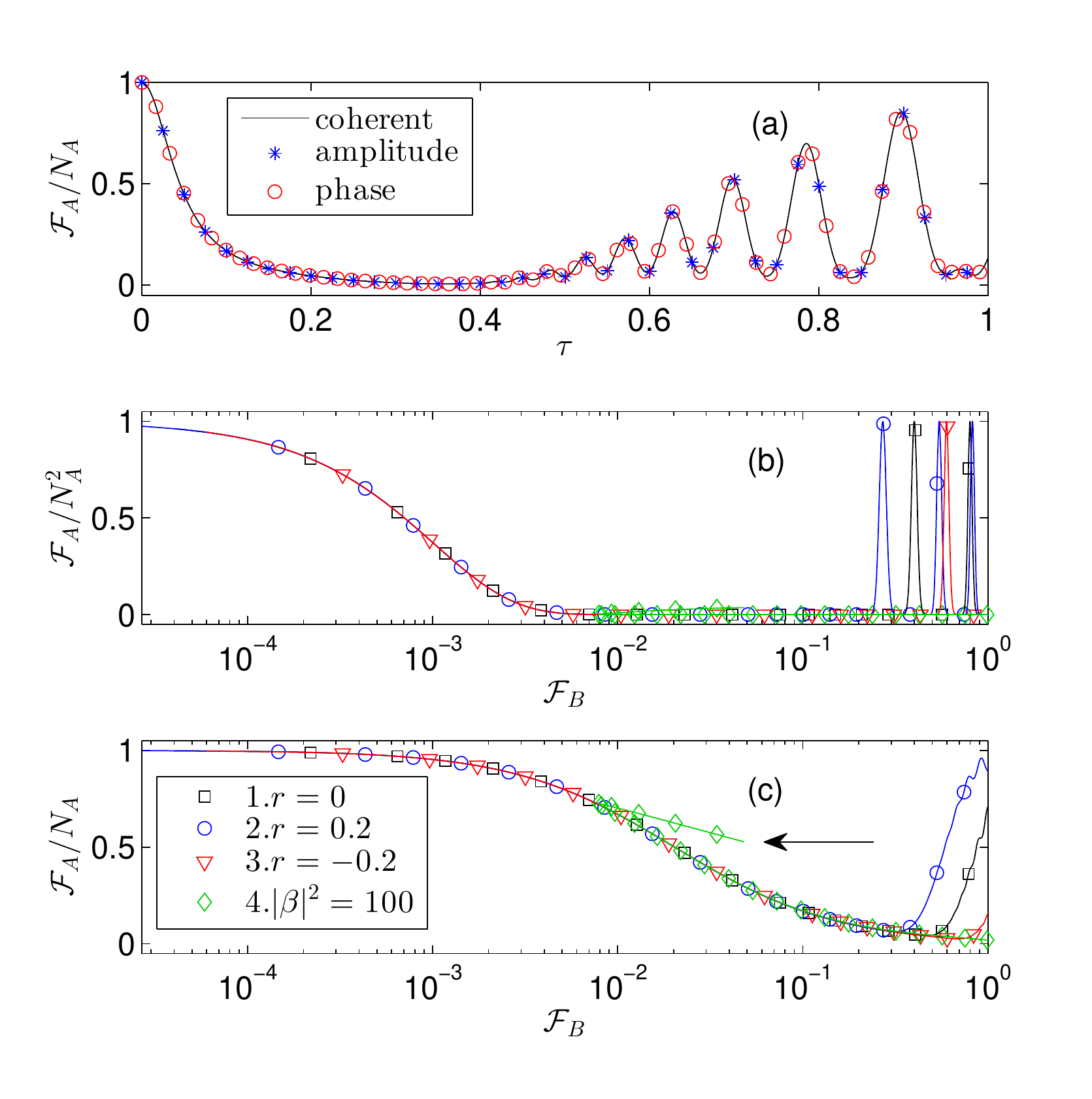} 
\caption{(Color online). Time evolution under $\hat{\mathcal{H}}_{AB} = \hbar g \hat{J}_z \otimes \hat{n}_B$ of the quantum Fisher information $\mathcal{F}_A$ and purity $\gamma$ of $\hat{\rho}_A(\tau)$, with $N_A=20$. In (a) we plot the QFI for the CSS $|\psi_A(0) \rangle = |\pi/2,0 \rangle$ for three different initial states, all with $\mathcal{F}_B=100$. We evolve from different auxiliary states, a Glauber-coherent state with $\beta = \sqrt{25}$, (solid black), an amplitude squeezed state with $\beta=\sqrt{50}$, $r \approx 0.352$ (blue asterisk), and a phase squeezed state with $\beta=\sqrt{20}$, $r \approx -0.111$  (red circles). The excellent agreement indicates that $\mathcal{F}_B$, rather than $|\psi_B(0)\rangle$ is the relevant quantity to consider when predicting this kind of decoherence, not only for $NOON$ states. In (b) and (c) we plot $\mathcal{F}_A$ against $\mathcal{F}_B$ for a $NOON$ state (b), normalized to $N_A^2$, and CSS (c), normalized to $N_A$. The QFI of the auxiliary system is varied in different ways for a number of initial states, (1-3) varying $|\beta|^2$ and (4) varying $r$, both with $\mathrm{arg}(\beta)=0$. Each point was evolved for a fixed rotation angle $\phi=\tau N_B(\beta, r)=\pi$ and as such in (b),(c) we take $\mathcal{F}_B=4V(\hat{n}_B/N_B)$. Arrow indicates over-squeezed regime.} 
\label{fig:varyFB}
\end{figure}

Evaluating the coherence matrix analytically for a squeezed coherent input state allows us to extend the short time results presented in the previous section to longer times. For $|\psi_A(0) \rangle = |NOON \rangle$ with generator $\hat{G}_B=\hat{n}_B$ [not, for instance $\hat{G}= \hat{n}_B/N_B$ which we consider in Fig.~\ref{fig:varyFB} (b),(c)], when $\beta^2 >> \sinh^2(r)$ we obtain 
\begin{equation} \label{eq:FAcoh}
\mathcal{F}_A \approx N_A^2 \exp \left(\frac{1}{2} \mathcal{F}_B \left[ \cos \left(N_A \tau \right)-1 \right] \right).
\end{equation}
Expanding this to second order in $N_A \tau$ recovers Eq.~\eqref{eq:FAapprox}, but this expression also predicts revivals in the QFI. Because this result was derived from $\mathcal{C}_{m-n}(\beta, r, \tau)$, we emphasize that unlike Eq.~\eqref{eq:FAapprox} it is not general in $|\psi_B(0) \rangle$, it only holds for squeezed coherent states and Fock states, the latter simply because $\mathcal{F}_B=0$. 

In Fig.~\ref{fig:varyt} we show the probe QFI and purity for a $NOON$ state compared to a coherent spin-state, for a number of input states, and clearly see the QFI of the auxiliary state correctly predicts the rate of decoherence. It is evident that coherent spin-states are more robust to this kind of decoherence. As we have shown, the QFI of $NOON$ states decay exponentially at a rate directly proportional to $N_A^2$, which is clearly not the case for coherent spin-states [see Eq.~\eqref{eq:purityapprox} and Eq.~\eqref{eq:FAapprox}]. As an example, using the experimental parameters of the system demonstrated in \cite{Zhang2012}, for the situation considered in Fig.~\ref{fig:varyt} with coherent light, the QFI of the NOON state would halve in approximately $10^{-6}$ seconds, while the QFI of the coherent spin state would take roughly an order of magnitude longer to decay by the same amount. Other time scales are discussed in the Figure legend.

In Fig.~\ref{fig:varyFB} we demonstrate that $\mathcal{F}_B$ remains an excellent predictor of decoherence where simple analytic expressions are unavailable. In Fig.~\ref{fig:varyFB} (a) we plot the QFI as a function of time for a CSS for three different input states, resulting in identical dynamics even for large times. This seems to indicate that our results are not restricted to spin-cat states. Up until now we have neglected the contribution of $\hat{n}_B$ to the magnitude of the rotation, i.e. to rotate the state about $\hat{J}_z$ by some angle $\phi$ we require an interaction time $\tau = \phi/N_B$. This must be taken into account in order to meaningfully compare the ability of different states $|\psi_B(0) \rangle$ to perform some fixed rotation $\phi$, so in Fig.~\ref{fig:varyFB} (b), (c) we take our generator to be $\hat{G}_B=\hat{n}_B/N_B$. In this case our full expression for $\mathcal{F}_A$ [Eq.~\eqref{eq:FAcoh}] does not hold, and although it is straight-forward to obtain a more general expression from the coherence matrix, it is not particularly enlightening. Because $\mathcal{F}_B(r)$ is not one-to-one, when the state is over squeezed there is a turning point in Fig.~\ref{fig:varyFB} (b), (c). We also see revivals which are predicted by Eq.~\eqref{eq:FAcoh}. These revivals occur as a result of the quantisation of the fields, for instance, if $|\psi_A(0) \rangle = |NOON \rangle $ they will occur when $\tau=2 \pi k/N_A$ where $k=1,2,3 ...$. 

Within the limitations discussed above, $\mathcal{F}_B$ entirely determines the subsequent dynamics. The starting point for this entire analysis was identifying an operator $\hat{G}_B$, which we were able to do because the reduced density matrix could be written in terms of $\mathcal{C}_{m-n}$ [Eq.~\eqref{eq:rhoA}], which was a direct consequence of the operator product form of the Hamiltonian. We now turn our attention to a beam-splitter Hamiltonian, where this is not the case.

\section{Non-Separable Interactions (Beam-Splitter)} \label{sec4}

A kind of interaction highly relevant to quantum metrology is an atomic beam-splitter; a non-photon conserving process that transfers population between our atomic modes. A common method for atomic interferometry is a Mach-Zehnder interferometer, which may be realised by performing two of these pulses, separated by a phase shift. This kind of evolution is also highly relevant to the preparation of spin-cat states, for instance it may be useful to rotate a $\hat{J}_y$ spin-cat state, perhaps generated via the atomic Kerr effect, to a $NOON$ state, which would require a rotation about $\hat{J}_x$. If we perform this rotation without assuming classical light, how might this decohere our atomic system? 

In particular, as the transfer of an atom is correlated with the creation or annihilation of a photon, the number of photons in the optical beam carries information about the number of transferred atoms, thus destroying the coherence of the superposition. This will be particularly relevant when creating $NOON$ states, as the creation of this state results in the creation of $ \sim \pm N_A/2$ photons which, depending on the initial state, may be easily distinguishable. If the Hamiltonian for this process is not separable, i.e of the form Eq.~\eqref{eq:sepham}, can we identify a generator and corresponding Fisher information which is a useful predictor for this decoherence?

\subsection{The Tavis-Cummings Model} \label{sec4a}

The fully quantized Hamiltonian for an atomic beam-splitter generated from atom-light interaction is the Tavis-Cummings Hamiltonian, which describes an ensemble of $N_A$, two-level atoms (with energy difference $\hbar \omega_0 = E_2-E_1$) interacting with a single mode optical field of frequency $\omega$ through dipole coupling \cite{Cummings1968}, 
\begin{align} \label{eq:TCham}
\mathcal{H}_{AB} =& \hbar \omega \hat{n}_B + \frac{1}{2}\hbar \omega_0 \hat{J}_z + \nonumber \\ 
&+\frac{1}{2}\hbar g \left(\hat{J}_+ + \hat{J}_- \right) \otimes \left( \hat{b}^\dagger+ \hat{b}\right).
\end{align}
If we were to ignore the quantum degrees of freedom of the light, the interaction term would simply result in a rotation about $\hat{J}_x = \frac{1}{2}(\hat{J}_+ + \hat{J}_-)$. 

In typical experimental systems the field is close to resonance, $\omega \approx \omega_0$ and the coupling $g$ is small compared to $\omega_0$, $\omega$. Therefore the rotating wave approximation is often made, and it is a good approximation to neglect the energy non-conserving terms $\hat{J}_+ \hat{b}^\dagger$ and $\hat{J}_- \hat{b}$. 

Before throwing away these terms, the interaction part of the Hamiltonian can be written as $\mathcal{H}_{\mathrm{int}} = \hbar g \hat{J}_x \otimes (\hat{b}+\hat{b}^\dagger)$ which certainly \emph{looks} separable, however the evolution caused by $\hat{\mathcal{H}}_0=\hbar \omega \hat{n}_B + \frac{1}{2}\hbar \omega_0 \hat{J}_z $ cannot be neglected. Moving into the interaction picture allows us to evolve the initial state forward in time under $\hat{\mathcal{H}}_{\mathrm{int}}$ only, but transforming this Hamiltonian into the interaction picture, and integrating the resultant interaction picture Hamiltonian in time gives rise to non-separable evolution. 

\begin{figure*}
\includegraphics[width=\textwidth]{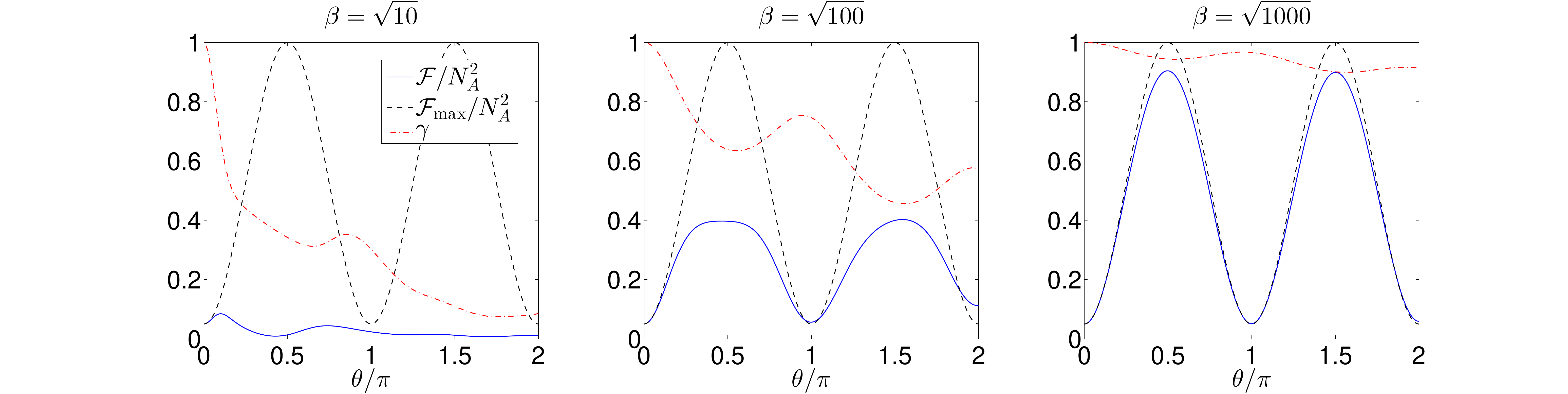}
\caption{(Color online). Loss of QFI and purity from evolution fully quantised beam-splitter Hamiltonian $\hat{\mathcal{H}}_{AB}= \hbar g ( \hat{X} \otimes \hat{J}_x+\hat{Y} \otimes \hat{J}_y )$ as a function of time. Time is normalized to the beam-splitter angle $\theta=2|\beta|\tau$, for $|\psi_A(0) \rangle = |NOON \rangle$ with $N_A=20$ and $|\psi_B(0) \rangle = |\beta \rangle$, with $\mathrm{arg}(\beta)=0$. The QFI is calculated with respect to $\hat{G}_A=\hat{J}_z$, such that after a $\pi/2$ rotation about the $\hat{J}_x$ axis the state approaches the QFI for a NOON state. $\mathcal{F}_{\mathrm{max}}$ is the QFI of the probe system in the limit of classical light, i.e. as the coherent amplitude becomes sufficiently large that we may substitute $\hat{b} \rightarrow \beta$ in the beam-splitter Hamiltonian [Eq.~\eqref{eq:Hint}].}
\label{fig:FAvst}
\end{figure*}

However, moving into the interaction picture reveals that quantities such as the purity of the reduced density matrix, and expectation values of any observable that commutes with $\hat{J}_z$ (such as the QFI with $\hat{G}_A=\hat{J}_z$) are unchanged by evolution under $\hat{\mathcal{H}}_0$. So long as we are only interested in calculating these quantities, we neglect $\hat{\mathcal{H}}_0$ and the interaction and Schr{\"o}dinger pictures coincide with
\begin{equation} \label{eq:Hint}
\hat{\mathcal{H}}_{AB}= \hbar g \left( \hat{X} \otimes \hat{J}_x+\hat{Y} \otimes \hat{J}_y \right)
\end{equation}
which we call the beam-splitter Hamiltonian, where $\hat{X}=\hat{b}+\hat{b}^\dagger$ and $\hat{Y}=-i(\hat{b}-\hat{b}^\dagger)$ are the standard optical amplitude and phase quadratures. To arrive at this Hamiltonian we have assumed the field is on resonance $\omega=\omega_0$ and have made the rotating wave approximation. 

As in Section \ref{sec2}, retaining a quantized description of the auxiliary system introduces decoherence to the evolution. Fig.~\ref{fig:FAvst} shows the QFI and the purity of a $\hat{J}_y$ spin-cat state ($|SC \rangle$) being rotated by a quantized beam-splitter [Eq.~\eqref{eq:Hint}] against the evolution time, parameterized by the beam-splitter angle $\theta=2|\beta| \tau$, compared to evolution under a classical beam-splitter $\hat{U}_{\mathrm{classical}}=\exp(-i \hat{J}_x \theta)$, obtained by taking the classical limit for the optical field $\hat{b} \rightarrow \beta$. We see that as $|\beta|^2$ becomes large, the full evolution approximates a classical beam-splitter.

\subsection{Identifying a Generator} \label{sec4b}
In Section \ref{sec3} we found that the QFI of the generator of time-evolution for system $B$ was an excellent tool for predicting decoherence. However, the difficulty with using this approach for decoherence introduced under the beam-splitter Hamiltonian [Eq.~\eqref{eq:Hint}], is that because the evolution is not separable, the reduced density matrix cannot be written in the form of Eq.~\eqref{eq:rhoA}. This means it is unclear how to identify a generator for the auxiliary system. Clearly under the beam-splitter Hamiltonian the optical field quadratures $\hat{X}$ and $\hat{Y}$ are responsible for generating the atom-light entanglement, however the basis in which the off-diagonal density matrix elements will decay depends on the argument of the coherent amplitude $\beta$. To isolate the role of the quantum fluctuations in each quadrature, we make the approximation that quantum fluctuations in one of the quadratures is negligible. Specifically, restricting ourselves to light with real coherent amplitude $\beta$, we compare the full quantum evolution [Eq.~\eqref{eq:Hint}] to two cases:
\begin{itemize}
\item Classical $Y$: $\hat{U}_X=\exp \left[ -i\left( \hat{X} \otimes \hat{J}_x + \langle \hat{Y} \rangle \hat{J}_y \right) \tau \right]=\exp\left[ -i\hat{X} \otimes \hat{J}_x \tau \right] $
\item Classical $X$: $\hat{U}_Y=\exp \left[ -i\left( \langle \hat{X} \rangle \hat{J}_x + \hat{Y} \otimes \hat{J}_y \right) \tau \right]$
\end{itemize} 
i.e. Eq.~\eqref{eq:Hint} with the substitution $\hat{Y} \rightarrow \langle \hat{Y} \rangle = 2\mathrm{Im}(\beta)=0$ for $\hat{U}_X$ and $\hat{X} \rightarrow \langle \hat{X} \rangle = 2 \beta $  for $\hat{U}_Y$. 

\begin{figure}
\includegraphics[width=\columnwidth]{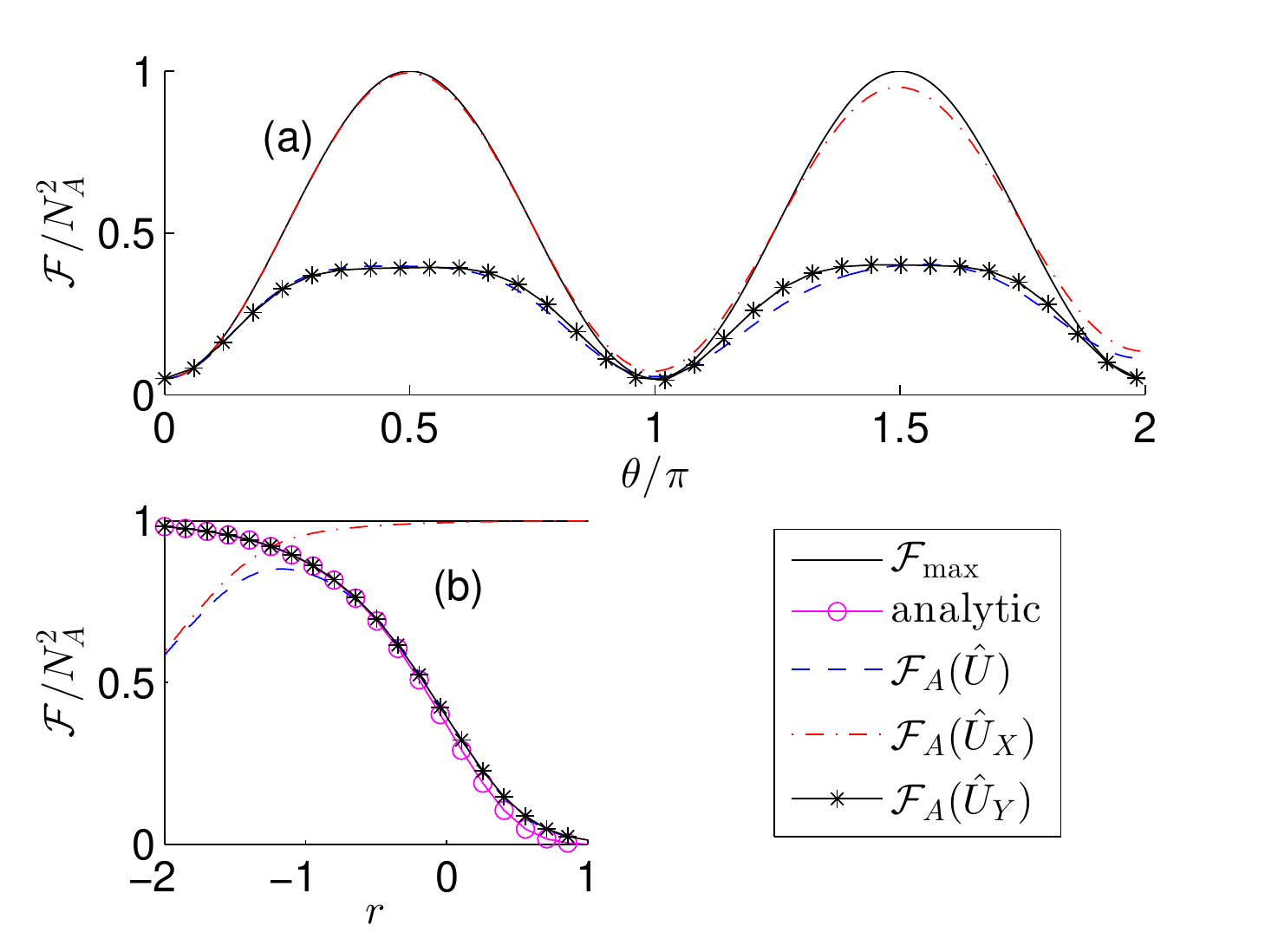} 
\caption{(Color online). Comparison of quantised beam-splitter evolution, neglecting quantisation of different optical quadratures. (a) QFI of probe system as a function of beam-splitter angle $\theta=2|\beta| \tau$ for each case with $|\psi_B(0) \rangle = |\beta \rangle$, and bottom: QFI as a function of squeezing magnitude. The system was simulated exactly for $N_A=20$ atoms and $\beta=\sqrt{100}$ for an initial $\hat{J}_y$ spin-cat state, rotated about the $\hat{J}_x$ axis. Both plots are a comparison of the fully quantized evolution [Eq.~\eqref{eq:Hint}] to the two cases ``classical X" and ``classical Y".  As in Fig.~\ref{fig:FAvst}, $\mathcal{F}_{\mathrm{max}}$ is the QFI due to the rotation only, without decoherence. In (b) we fix $\theta=\pi/2$ (the optimum value) and vary $r$. The magenta circles are an analytic calculation [Eq.~\eqref{eq:FABS}], based off the approximate generator $\hat{G}_B \approx \hat{Y}/\langle \hat{X} \rangle$.}
\label{fig:casevscase}
\end{figure}

Fig.~\ref{fig:casevscase} shows the QFI of a $\hat{J}_y$ spin-cat state evolved under Eq.~\eqref{eq:Hint} compared to the two cases $\hat{U}_X$ and $\hat{U}_Y$. For comparison, we have also shown $\mathcal{F}_{\mathrm{max}}$. This is the QFI $\mathcal{F}_A= 4V(\hat{J}_z)$ for a $\hat{J}_y$ spin-cat state evolved under $\hat{U}_{\mathrm{classical}}=\exp(-i \hat{J}_x \theta)$ which imparts no decoherence, therefore $\mathcal{F}_A < \mathcal{F}_{\mathrm{max}}$. Fig.~\ref{fig:casevscase} (a) indicates that $\hat{U}_X$ agrees well with the classical evolution and has only a small impact on the coherence, and that $\hat{U}_Y$ agrees well with the evolution due to Eq.~\eqref{eq:Hint}. Fig.~\ref{fig:casevscase} (b) varies the squeezing parameter $r$ at the optimum beam-splitter angle $\theta=2|\beta| \tau = \pi/2$, and shows good agreement with the outcome of Fig.~\ref{fig:casevscase} (a) for moderate $|r|$, i.e. that fluctuations in $\hat{Y}$ are predominantly responsible for the decoherence. This picture breaks down for highly phase squeezed initial auxiliary states, and it becomes important to consider quantum fluctuations in $\hat{X}$ rather than $\hat{Y}$ to correctly describe the system. 

Motivated by Fig.~\ref{fig:casevscase} we continue by studying evolution under $\hat{U}_Y$ only. Ignoring the quantum fluctuations in $\hat{X}$ allows us to define the commuting operators $\hat{\phi}=\arctan(\hat{Y}/\langle \hat{X} \rangle)$ and $\hat{\theta}=\langle \hat{X} \rangle \tau \sqrt{1+(\hat{Y}/\langle \hat{X} \rangle)^2}$, such that
\begin{equation}
e^{-i(\langle \hat{X} \rangle \hat{J}_x + \hat{Y}\hat{J}_y)\tau} = e^{-i \hat{\phi} \hat{J}_z} e^{-i \hat{\theta} \hat{J}_x} e^{i \hat{\phi} \hat{J}_z}. \label{SimonR}
\end{equation}
Evaluating expectation values with respect to a squeezed coherent state gives $\langle \hat{X} \rangle = 2 |\beta| \approx 2 \sqrt{N_B}$ if $|\beta|^2>>\sinh^2(r)$. Thus for large $N_B$ we expand in $1/\langle \hat{X} \rangle$ to first order, giving $\hat{\phi} \approx \hat{Y}/\langle \hat{X} \rangle$ and $\hat{\theta} \approx \langle \hat{X} \rangle \tau \equiv \theta$. This decouples the part of the Hamiltonian which generates entanglement from the part that generates the rotation. 

Restricting ourselves to an initial state $|\psi_A(0)\rangle = |SC\rangle$, and $\theta = \pi/2$, we note that $\exp (-i \hat{\phi}\hat{J}_z)$ will cause dephasing of off-diagonal terms in the $J_z$ basis. The $\exp (-i\theta \hat{J}_x)$ term will then rotate this state such that it is approximately aligned with the maximal and minimal $J_z$ eigenstates, before it undergoes further decoherence due to $\exp (-i \hat{\phi}\hat{J}_z)$. For $N_A \gg 1$, if the fluctuations in $\hat{\phi}$ are much less than $1$, then this second dephasing process will be much more significant than the first, as the off-diagonal terms are significantly more separated after the $\pi/2$ rotation. As such, it is a reasonable approximation to neglect the effect of the first dephasing step, and in terms of the pseudo-spin eigenspectrum $\hat{J}_\alpha |m; \alpha \rangle = \lambda_m^\alpha |m; \alpha \rangle$ with $\alpha=x,y,z$, the reduced density matrix of system $A$ becomes
\begin{align} \label{eq:rhoABS}
\hat{\rho}_A(\theta) \approx & \sum_{m,m', n, n'} c_{m'} (c_{n'})^* A_{m,m'} (A_{n,n'})^* \mathcal{C}_{m-n}^z \times  \\ \nonumber 
& \times |m; z \rangle \langle n; z|e^{-i \left(\lambda_{m'}^x-\lambda_{n'}^x \right) \theta}
\end{align}
where $c_j=\langle j; x| \psi_A(0) \rangle$ is this initial sate in the $\hat{J}_x$ eigenbasis and $A_{j,k} = \langle j; z | k; x \rangle$ is a change of basis. In analogy to Eq.~\eqref{eq:coh} we identify (using $\lambda_j^z = j$)
\begin{equation} \label{eq:cohBS}
\mathcal{C}_{m-n}^z= \langle \psi_B (0)| e^{-i \left( m - n \right) \hat{Y}/\langle \hat{X} \rangle  } | \psi_B (0)\rangle
\end{equation}
as the term responsible for decay of coherence in the $\hat{J}_z$ eigenbasis. 

As an example, for Glauber-coherent states (with $\beta$ real)
\begin{equation}
\mathcal{C}_{m-n}^z(\beta)=\exp\left[-\frac{(m-n)^2}{8N_B}\right]
\end{equation} 
Now, proceeding as in Section \ref{sec3} we can identify the QFI for the auxiliary system as $\mathcal{F}_B=4V(\hat{G}_B)$ with
\begin{equation} \label{eq:GB}
\hat{G}_B \approx \hat{Y}/\langle \hat{X} \rangle. 
\end{equation}
For coherent states, $V(\hat{Y}) =1$, so $\mathcal{F}_B = 1/N_B$, indicating that increasing the number of photons used to implement the beamsplitter reduces the decoherence. A qualitative explanation for this is, if the initial state contains a large number of photons, it is more difficult to distinguish the creation or annihilation of $\sim N_A/2$ photons. Conversely, a Fock state has $V(\hat{Y}) = 2 N_B +1$, and $\langle \hat{X}\rangle =0$, indicating that it has a very high QFI and will cause extremely rapid decoherence. Again, this fits with our intuitive picture, as the creation or annihilation of one photon from a Fock state is immediately distinguishable, indicating that it cannot be used to create a coherent superposition of atomic population. The generator $\hat{G}_B \approx \hat{Y}/\langle \hat{X} \rangle$ also indicates that phase squeezed states should cause less decoherence than amplitude squeezed states. 

If we are restricted to rotations $\theta=\pi/2$, such that $\exp \left(-i \hat{J}_x \pi/2 \right)|SC \rangle = |NOON\rangle$, then the reduced density matrix takes the form of Eq.~\eqref{eq:rhoA}, and the results obtained in Section \ref{sec3} can be applied here but with $\hat{G}_B=\hat{Y}/\langle \hat{X} \rangle$. We have
\begin{equation} \label{eq:FABS}
\mathcal{F}_A \left(\theta=\frac{\pi}{2} \right) \approx N_A^2 \exp \left(-\frac{1}{4} \mathcal{F}_B N_A^2 \right),
\end{equation}
which although similar to Eq.~\eqref{eq:FAapprox}, does not depend on time as we have fixed $\theta$. The purity can be obtained from Eq.~\eqref{eq:analNoonQFI}. This result agrees well to the exact (numeric) evolution, shown in Fig.~\ref{fig:casevscase}, indicating that the generator is well approximated by $\hat{G}_B \approx \hat{Y}/\langle \hat{X} \rangle$.

\section{A Semi-Classical Picture of Decoherence} \label{sec5}

The results presented in the previous section were obtained by evolving the full quantum state $|\psi_{AB}\rangle$, which becomes increasing challenging as our basis size increases. We also found that it was an excellent approximation in most regimes to neglect the quantum fluctuations in one quadrature, which allowed us to treat the interaction as separable such that we could identify a generator for system $B$. Here we present an approximate, general approach to studying decoherence arising from the entanglement of a probe with an auxiliary quantum field. This approach does not require us to neglect quantum fluctuations in $\hat{X}$ or $\hat{Y}$ to identify a generator, and also affords us an efficient way of simulating the system numerically. We make use of this in Section \ref{sec6}, where we use this method to explore the required auxiliary field occupation to negate decoherence arising from the entanglement between the systems.

This is done by modelling the reduced density matrix as an average over a set of noisy classical variables $\mathbf{X}$, which have some distribution function $P(\mathbf{X})$, characterised by $|\langle \mathbf{X}|\psi_B(0) \rangle|^2$. Following a series of measurements, the reduced density matrix is well approximated by
\begin{equation} \label{eq:rhoapprox}
\hat{\rho}_A \approx \int d\mathbf{X} P(\textbf{X}) \hat{U}(\mathbf{X})|\psi_A (0) \rangle \langle \psi_A(0)| \hat{U}^\dagger (\mathbf{X}).
\end{equation}
A similar model of decoherence has been considered elsewhere, where it was used to prove a general link between the probe QFI and purity \cite{Modi2016}. This relation is approximate in the sense that the quantisation of the optical field is neglected, for instance revivals predicted by Eq.~\eqref{eq:FAcoh} are absent in this picture. Nevertheless, for sufficiently short times, Eq.~\eqref{eq:rhoapprox} is an excellent approximation to the exact dynamics followed by a partial trace. In the inset of Fig.~\ref{fig:NbvsNa} we compare this method to an exact calculation for small $N_A$ for the beam-splitter Hamiltonian, and find excellent agreement.

Although conceptually similar, we emphasise that this approach is distinct from stochastic phase space methods \cite{maviswigner, Olsen2009} commonly used to model Bose-Einstein condensates beyond a mean-field treatment, such as the well known truncated Wigner approximation \cite{Drummond1993, Werner1995, Sinatra2001}. Significantly, in these phase space methods expectation values of observable quantities are reconstructed by averaging over phase space trajectories, whereas in this method we have full access to the (approximate) reduced density matrix. This allows us to easily calculate the QFI, which for a mixed state, is difficult to obtain via a phase space method. Additionally, although we often evaluate the integral in Eq.~\eqref{eq:rhoapprox} numerically, we do this by performing a Riemann sum over $P(\textbf{X})$ rather than stochastically sampling from the distribution.

For the two rotations we study, it is useful to choose these noisy, classical variables to be the Bloch sphere angles $\mathbf{X}=\{ \phi \}$ for the separable Hamiltonian or $\mathbf{X}=\{ \theta, \phi \}$ for the beam-splitter Hamiltonian. This approach has a number uses, for instance it is simple in this picture to study the effects of entanglement generated by $\hat{X}$ and $\hat{Y}$ simultaneously. Additionally, it is only ever necessary to manipulate matrices which belong to the probe vector space, rather than constructing and evolving the full $\dim(A) \times \dim(B)$ state before performing a partial trace to obtain $\hat{\rho}_A$, which rapidly becomes intractable even for modest particle numbers.

\subsection{$\hat{J}_z$ Rotation} \label{sec5a}

As an example we first show that this method can recover the results presented in Section \ref{sec3}. As we have alluded to, decoherence under the separable Hamiltonian [Eq.~\eqref{eq:Jzrot}] can be understood by averaging over a single parameter $\mathbf{X}=\{\phi\}$ with $\hat{U}(\phi)=\exp(-i \hat{J}_z \phi)$, with the noise properties of $\phi$ related to the quantum fluctuations of the operator $\hat{\phi} = \hat{n}_B \tau$. In the $\hat{J}_z$ eigenbasis this gives the reduced density matrix
\begin{equation} \label{eq:rhoapproxJz}
\hat{\rho}_A \approx \sum_{m,n} c_m c_n^* \int d\phi P(\phi) e^{-i \left( m-n\right) \phi} |m \rangle \langle n | .
\end{equation}
If we identify $\mathcal{C}_{m-n}=\int d\phi P(\phi) e^{-i \left( m-n\right) \phi}$, this has the same form as Eq.~\eqref{eq:rhoA}. If we assume $P(\phi)$ is Gaussian with mean $\overline{\phi}$ and standard deviation $\sigma_{\phi}$, then we can evaluate this integral to obtain 
\begin{equation} \label{eq:noisycoh}
\mathcal{C}_{m,n}^z=e^{-i \left(m- n \right) \overline{\phi}}e^{-\frac{1}{2}\left(m - n \right)^2 \sigma_{\phi}^2},
\end{equation}
adding the $z$ superscript to denote decoherence in the $\hat{J}_z$ eigenbasis. For coherent light, this expression agrees with Eq.~\eqref{eq:cohbeta} by identifying $\overline{\phi}=\langle \hat{\phi}\rangle  =|\beta|^2 \tau$ and $\sigma_{\phi}^2= V(\hat{\phi}) = |\beta|^2 \tau^2$, which is seen easily by expanding $\exp[-i (m - m )\tau]$ to second order in $\tau$. 

We have identified $\mathcal{F}_B=4V(\hat{n}_B)$ which tells us that $\hat{G}_B=\hat{n}_B$ and so we associate $\phi$ with the mean and noise properties of the operator $\hat{n}_B$. Although this description correctly predicts $\hat{G}_B$ it is only approximate, and because we have neglected the quantisation of the photon field it will not capture the revivals seen in Fig.~\ref{fig:varyt} (c) or (d).

\subsection{Beam-Splitter} \label{sec5b}

Now we turn our attention to studying the decoherence generated by evolution under the beam-splitter Hamiltonian [Eq.~\eqref{eq:Hint}]. As in Section \ref{sec4b} we study the approximate rotation $\hat{U}(\theta,\phi) = \exp(-i \hat{J}_z \phi) \exp(-i \hat{J}_x \theta)$, identifying $\mathbf{X}=\{\theta,\phi\}$. Again, we assume the distribution functions for $\theta$ and $\phi$, $P(\theta, \phi)=Q(\theta)Q(\phi)$ are Gaussian. We interpret $(\overline{\theta}, \overline{\phi})$ as the azimuthal and elevation Bloch sphere angles respectively, and identify
\begin{align} \label{eq:XProt}
X&=\frac{\theta \cos(\phi)}{\tau}\\
Y&=\frac{\theta \sin(\phi)}{\tau} \nonumber
\end{align}
These are to be interpreted as noisy classical variables with $\overline{X}=\langle \hat{X} \rangle$ and $\sigma_X^2 = V(\hat{X})$ (with analogous relations for $Y$), which imply that the angles are related to the coherent amplitude of the light, with $\overline{\theta}=2|\beta|\tau$ and $\overline{\phi}=\arg(\beta)$. 

Following a procedure similar to that in Section \ref{sec5a} we arrive at the reduced density matrix,
\begin{align} \label{eq:nonseprho}
\hat{\rho}_A \approx \sum_{m,m',n,n'} &c^x_{m'} (c^x_{n'})^* A_{m,m'}(A_{n,n'})^*  \\ \nonumber
&\times  \mathcal{C}^z_{m,n}(\overline{\phi}, \sigma_{\phi}) \mathcal{C}^x_{m',n'}(\overline{\theta}, \sigma_{\theta}) |m, z \rangle \langle n, z |,
\end{align}
which is of form of Eq.~\eqref{eq:rhoABS}, with the difference that the phase factor has been replaced by $\mathcal{C}_{m',n'}^x$ which directly causes decay of the off-diagonal matrix elements in the $\hat{J}_x$ eigenbasis also. Both $\mathcal{C}_{j,k}^{x}$, $\mathcal{C}_{j,k}^{z}$ have the same form as Eq.~\eqref{eq:noisycoh}, but in terms of the relevant classical variable. 

The reduced density matrix Eq.~\eqref{eq:nonseprho} with the relations Eq.~\eqref{eq:XProt} afford us an understanding of decoherence in terms of the noise properties of the optical quadratures $\hat{X}$ and $\hat{Y}$. If $\beta$ is real, we set $\overline{\phi}=0$, (which corresponds to performing our rotations about $\hat{J}_x$ only) we obtain the following noise relations 
\begin{align} \label{eq:noiserlns}
\sigma_\theta^2 &= \frac{\overline{\theta}^2 \  V(\hat{X})}{4 |\beta|^2} \\
\sigma_\phi^2 &= \frac{ V(\hat{Y})}{4 |\beta|^2} \nonumber
\end{align}
Observing that $\langle \hat{X} \rangle = 2 |\beta|$, these relations agree with result Eq.~\eqref{eq:GB} that the generator responsible for decay of the off-diagonal matrix elements of $\hat{\rho}_A$ in the $\hat{J}_z$ eigenbasis is $\hat{G}_B^z \approx \hat{Y}/\langle \hat{X} \rangle$, but they also allow us to identify $\hat{G}_B^x \approx \overline{\theta} \hat{X}/\langle \hat{X} \rangle$, as the generator of decay in the $\hat{J}_x$ eigenbasis. However, Fig.~\ref{fig:casevscase} indicates that for the rotation of a $\hat{J}_y$ spin-cat state about $\hat{J}_x$, noise in $\hat{Y}$ dominates.

\section{Mitigating Decoherence} \label{sec6}

In Fig.~\ref{fig:FAvst}, it is apparent that as $\beta$ increases, $\mathcal{F}_A$ approaches the classical limit. This agrees with the result that $\mathcal{F}_B = 1/|\beta|^2$, as  $\hat{G}_B^z\approx \hat{Y}/\langle \hat{X} \rangle$, with $\langle \hat{X} \rangle = 2|\beta|$ and $V(\hat{Y}) = 1$ (for coherent light). We also see this behaviour in Fig.~\ref{fig:varyFB} (b) and (c), when comparing $\mathcal{F}_A$ for a fixed rotation angle $\phi=\tau N_B$ the decoherence vanishes as $\mathcal{F}_B=4V(\hat{n}_B)/N_B^2 \propto 1/N_B$ goes to zero, which corresponds to the limit of large photon number. Motivated by these observations, here we study the following question: given evolution under either of the entangling Hamiltonians we have considered, what is the required auxiliary field occupation to mitigate decoherence in the probe system?

More specifically, we calculate the required $N_B$, such that after rotating the probe state by a fixed angle the probe QFI has at least $\mathcal{F}_A=N_A^2/2$, which we will call $N_B^{TFS}$. This is the QFI of the twin-Fock state, defined $|TFS \rangle = |N_A/2, N_A/2 \rangle$ with respect to $\hat{G}_A=\hat{J}_y$. Our motivation for this metric is that twin-Fock states are far less exotic than spin-cat states, and can be realized simply by a projective measurement in the $J_z$ basis. Superpositions of twin-Fock states also have $\mathcal{F}_A \approx N_A^2/2$, and can be manufactured via any pair-wise particle creation process, such as four-wave mixing \cite{Dall2009, Bucker2011} and spin-exchange collisions \cite{Lucke2011, Gross2011, Bookjans2011, Hamley2012}. Although a TFS would be less attractive than a $NOON$ state for a number of fundamental tests, it is an excellent candidate for quantum metrology. If one had a spin-cat state, and were unable to maintain the QFI above what could be achieved with a twin-Fock state (which is much simpler to create), it would be much less challenging to simply use the latter. 

\subsection{$\hat{J}_z$ Rotation}

As we have analytic results for rotating a $NOON$ state under $\mathcal{H}_{AB}=\hbar g \hat{J}_z \otimes \hat{n}_B$, this is our starting point. From Eq.~\eqref{eq:FAapprox}, making the substitution $\tau=\phi/N_B$ we obtain
\begin{equation} \label{eq:NBTFS}
N_B^{TFS} \left[|NOON \rangle \right] \approx \frac{\phi^2 e^{-2r}}{\log(2)} N_A^2,
\end{equation}
which is valid within the same approximations as Eq.~\eqref{eq:FAapprox}. Although this does not explicitly depend on $\mathcal{F}_B$, as expected states with larger $\mathcal{F}_B$ per photon (for instance phase squeezed states) would require more photons to perform this rotation while maintaining $\mathcal{F}_A \geq N_A^2/2$.

This scaling is intuitive if we consider that the information relating to the $\hat{J}_z$ projection of system $A$ is encoded onto $|\psi_B\rangle$ as a phase shift. In order to maintain coherence between the maximal and minimal $J_z$ eigenstates, we require this information be \emph{hidden} in the quantum fluctuations of  the phase of $|\psi_B(0)\rangle$. More specifically, in order to maintain indistinguishability, we require that the magnitude of the phase shift after time $\tau$, say $\phi_{\rm cat}=\phi N_A/N_B$, that each component of the superposition cause on $|\psi_B(0)\rangle$ is less than the characteristic phase fluctuations of $|\psi_B(0)\rangle$, $V(\hat{n}_B/N_B) \sim e^r/\beta$. Setting $\phi_{\rm cat} \sim \sqrt{V(\hat{n}_B/N_B)}$ gives $N_B^{TFS} \sim e^{-2r} \phi^2 N_A^2$. 

\begin{figure}
\includegraphics[width=\columnwidth]{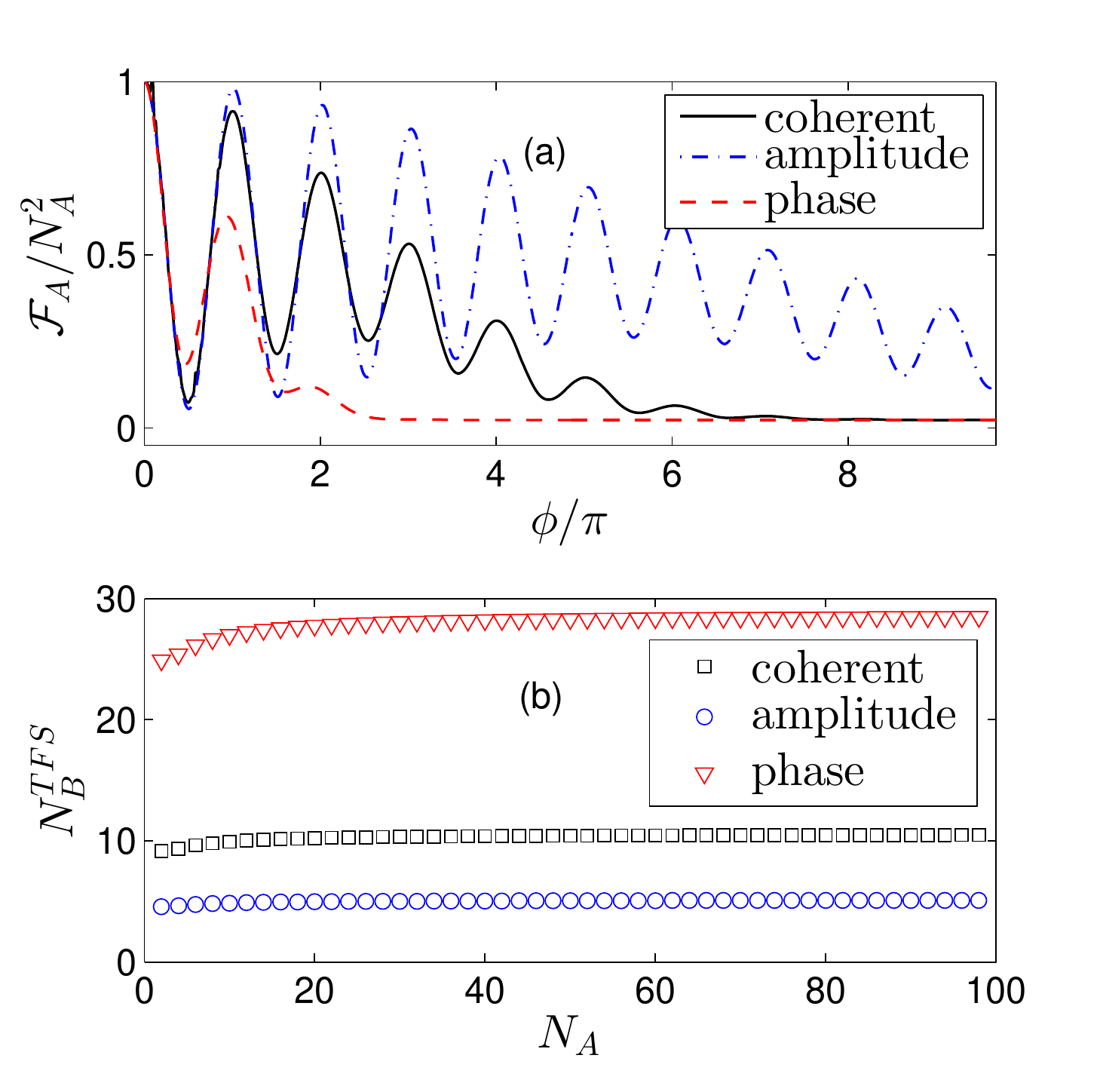} 
\caption{(Color online). (a) $\mathcal{F}_A$ for a $\hat{J}_y$ spin-cat state under a $\hat{J}_z$ rotation as a function of rotation angle $\phi=\tau N_B$. Simulation was performed exactly for $N_A=20$ and $N_B=100$, with $|r|=1$ for the amplitude and phase squeezed states. (b) $N_B^{TFS}$ after a $\phi=\pi$ rotation, which corresponds to the first revival in (a) for $r=|0.5|$.}
\label{fig:SC}
\end{figure}

Interestingly, $\hat{J}_y$ spin-cat states are surprisingly robust to decoherence arising from this Hamiltonian. Fig.~\ref{fig:SC} (a) plots the $\mathcal{F}_A$ for this state as a function of time (parameterized by the rotation angle $\phi=N_B \tau $), calculated with respect to $\hat{G}_A=\hat{J}_y$. The oscillations in $\mathcal{F}_A$ are a consequence of the rotation , $\mathcal{F}_A$ is maximum when the state is aligned along the $\hat{J}_y$ axis. In Fig.~\ref{fig:SC} (b) we plot $N_B^{TFS}$ for a $\phi=\pi$ rotation, which corresponds to the first revival in Fig.~\ref{fig:SC} (a). The quadratic scaling $N_B^{TFS} \propto N_A^2$ exhibited by $NOON$ states under this rotation is not evident here, instead we find that $N_B^{TFS} $ is approximately independent of $N_A$.

The origin of the $N_A^2$ scaling for $NOON$ states is the linear $N_A$ dependence of $\phi_{\rm cat}$, which is absent for a $\hat{J}_y$ spin-cat rotating about the $\hat{J}_z$ axis. Here, the coherence is carried by the distinguishability of extreme $\hat{J}_y$ eigenstates. Fluctuations in $\hat{G}_B = \hat{n}_B/N_B$ will cause diffusion of the phase of each branche of the superposition. This phase diffusion will be of order $\Delta \phi = \sqrt{V(G_B)} \sim e^{-r}/\beta$. As $\Delta\phi$ increases, the separation between the two branches decreases, becoming indistinguishable when $\Delta\phi \sim \pi/2$. In this case the non-classical nature of the state is lost, and we expect $F_A\sim N_A$. We expect that the phase diffusion that leads to $F_A = N_A^2/2$ will occur well before this at some value $\Delta \phi_{\rm TFS}$. Setting $\Delta \phi = \Delta \phi_{\rm TFS}$ gives
\begin{equation}
N_B^{TFS} \left[ |SC \rangle \right] \approx \left( \frac{\phi}{\Delta \phi_{TFS}} \right)^2 e^{-2r},
\end{equation}
and from Fig.~\ref{fig:SC} (b) we estimate $\Delta \phi_{TFS} \approx \pi/3$ is a good rule of thumb.

\subsection{Beam-splitter}

\begin{figure}
\includegraphics[width=\columnwidth]{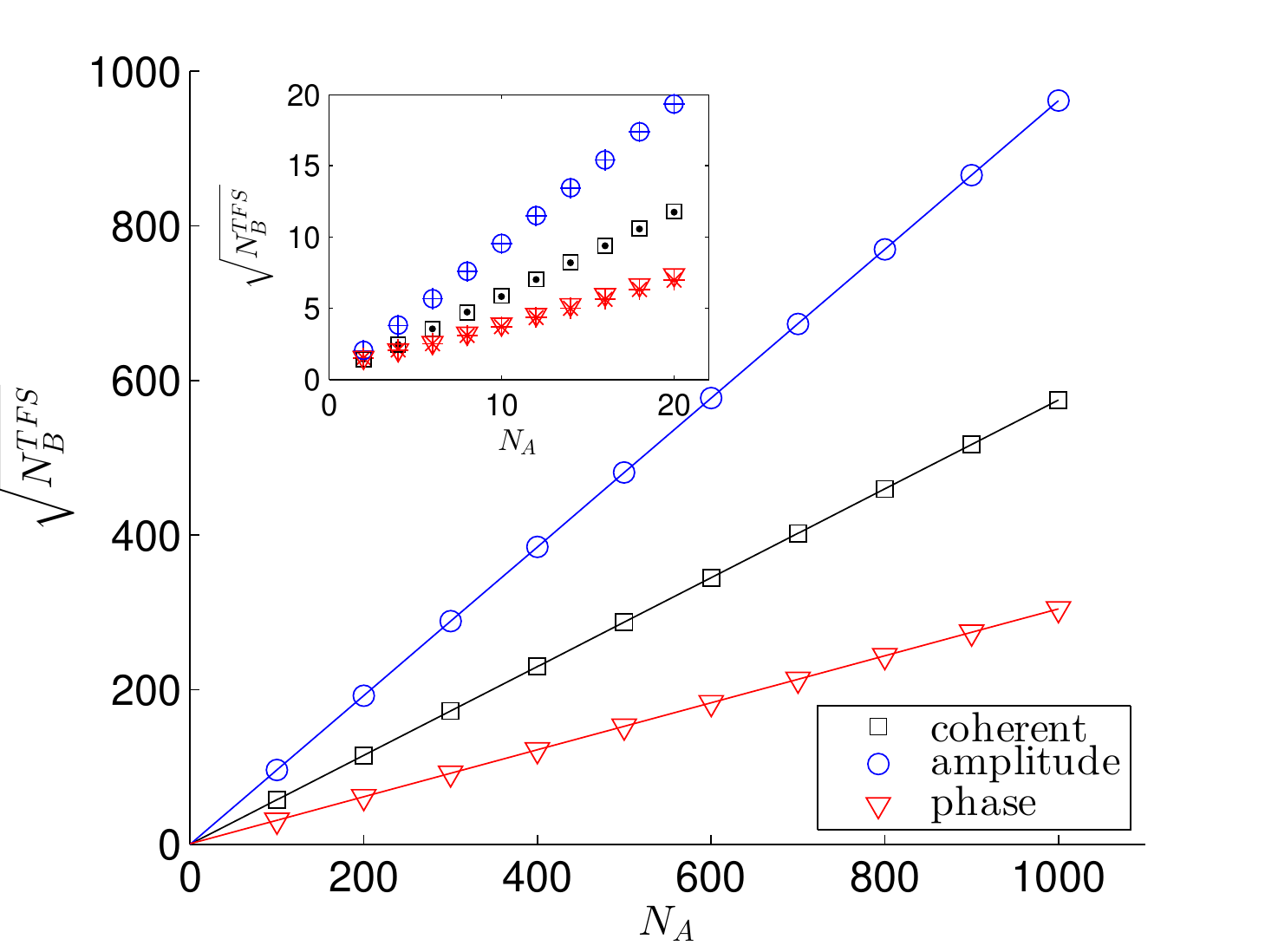} 
\caption{(Color online). The required number of photons such that the rotated spin-cat state has QFI equal to that of a Twin-Fock state, for a Glauber-coherent state, amplitude and phase squeezed states. Lines are a least-squares linear fit, the gradients are for comparison to Eq.~\eqref{eq:NBTFS2}. Simulation was performed with $\arg(\beta)=0$ and $r=0.5$ for the squeezed states, using the semi-classical picture [Eq.~\eqref{eq:rhoapprox}]. $N_B$ was varied by changing $|\beta|^2$ rather than $r$. Inset: Comparison of exact solution (points) to semi-classical picture (shapes) for small particle numbers.}
\label{fig:NbvsNa}
\end{figure}

Likewise, we can use Eq.~\eqref{eq:FABS} to estimate $N_B^{TFS}$ for a $\hat{J}_y$ spin-cat rotating about $\hat{J}_x$ by $\theta =\pi/2$, we obtain
\begin{equation} \label{eq:NBTFS2}
N_B^{TFS} \approx \frac{e^{2r}}{4 \log(2)}N_A^2.
\end{equation}
Crucially, in this relation the squeezing factor $\exp(2r)$ has the opposite sign to that of Eq.~\eqref{eq:NBTFS}, as we expect from $\mathcal{F}_B \approx 4 V(\hat{Y}/\langle \hat{X} \rangle)$ states with small fluctuations in $\hat{Y}$ will require the least number of photons. As this result is approximate we compare it to a numeric solution in Fig.~\ref{fig:NbvsNa}, both using exact diagonalisation for small $N_A$, and using the semi-classical picture presented in Section \ref{sec5} for a much larger range of $N_A$. We find excellent agreement between the exact numerics, the semi-classical picture and this analytic result, which uses the approximate generator $\hat{G}_B \approx \hat{Y}/\langle \hat{X} \rangle$.

\section{Conclusions} \label{sec7}

In quantum metrology it is sometimes necessary to prepare a state for input into a metrological device via an operation such as a beam-splitter or $\hat{J}_z$ rotation. This evolution may be performed via an interaction with an auxiliary system, and although it is commonplace to assume this auxiliary system is sufficiently large that any entanglement between the two systems may be neglected, here we retain a quantized description of both systems. We find that the QFI associated with the auxiliary system's ability to estimate the $\hat{J}_z$ projection of our primary system through the interaction Hamiltonian is an excellent predictor of decoherence and loss of metrological usefulness.

It is simple to define this QFI for a separable Hamiltonian [Eq.~\eqref{eq:sepham}], and we also derive an approximate QFI for a beam-splitter Hamiltonian [Eq.~\eqref{eq:Hint}]. By introducing an alternative picture of this decoherence, viewing the reduced density matrix as an average over an ensemble of noisy classical variables, we are also able to generalize our result for the beam-splitter case by defining two generators responsible for the decay of off-diagonal coherence in both the $\hat{J}_x$ and $\hat{J}_z$ eigenbases. In summary it is desirable to chose initial auxiliary states with small QFI, especially for $NOON$ states which are particularly susceptible to this kind of decoherence, see Fig.~\ref{fig:varyt}.

We have also estimated the required auxiliary field occupation to negate this kind of decoherence in both situations. As an example, it would require roughly $1 \times 10^5$ coherent photons to impart a $\pi$ phase shift on a 100 atom $NOON$ state, or about $3 \times 10^3$ coherent photons to rotate a 100 atom $\hat{J}_y$ spin-cat state by $\pi/2$ about the $\hat{J}_x$ axis, while maintaining the QFI above that of a twin-Fock state.

\begin{acknowledgements}

The authors would also like to thank Iulia Popa-Mateiu, Stuart Szigeti, Joel Corney, Michael Hush and Murray Olsen for invaluable discussion and feedback. The authors also made use of the University of Queensland School of Mathematics and Physics high performance computing cluster `Obelix', and thank Leslie Elliot and Ian Mortimer for computing support. SAH acknowledges the support of Australian Research Council Discovery Early Career Research Award DE130100575. This project has received funding from the European Union's Horizon 2020 research and innovation programme under the Marie Sklodowska-Curie grant agreement No 704672. 

\end{acknowledgements}

\bibliographystyle{apsrev4-1}

\bibliography{../../../nolan_quantmetrology_bib}

\end{document}